\documentclass[%
 reprint,
 amsmath,amssymb,
 aps,
]{revtex4-2}
\usepackage{caption}
\usepackage{float} 
\usepackage{graphicx}% Include figure files
\usepackage{dcolumn}% Align table columns on decimal point
\usepackage{bm}% bold math
\usepackage[colorlinks=true, linkcolor=blue, citecolor=blue, urlcolor=blue]{hyperref}% add hypertext capabilities
\begin{document}

\preprint{APS/123-QED}

\title{Low-exposure, high-quality multimodal speckle X-ray imaging via an intrinsic gradient-flow approach}% Force line breaks with \\

\author{Jayvan Liu} 
 \email{jli388@outlook.com}
\affiliation{School of Physical and Chemical Sciences, University of Canterbury, Christchurch, New Zealand
}%
\author{Samantha J. Alloo}%
 \email{samantha.alloo@monash.edu}
\affiliation{School of Physics and Astronomy, Monash University, Victoria 3800
}%

\author{Max Langer}
\email{max.langer@univ-grenoble-alpes.fr}
\affiliation{Univ. Grenoble Alpes, CNRS, UMR 5525, VetAgro Sup, Grenoble INP, TIMC, F-38000 Grenoble, France
}%
\author{Konstantin M. Pavlov}
 \email{konstantin.pavlov@canterbury.ac.nz}
\affiliation{School of Physical and Chemical Sciences, University of Canterbury, Christchurch, New Zealand \\
School of Physics and Astronomy, Monash University, Victoria 3800, Australia and \\ 
School of Science and Technology, University of New England, NSW 2351, Australia
}%

\date{\today}% It is always \today, today,

\begin{abstract}
We present a new approach for retrieving dark-field, phase shift, and attenuation images from speckle-based X-ray imaging data. Speckle-based X-ray imaging (SBXI) exploits sample-induced alterations to a reference near-field speckle pattern produced by a randomly structured mask. Attenuation images allow materials of different densities to be visualised. Phase-shift images are useful because they reveal how materials in a sample refract the X-ray beam, providing contrast between similar low-density structures that are difficult to reconstruct in attenuation images. Dark-field images convey information about structures that are smaller than the spatial resolution and thus invisible in both attenuation and phase-shift images. In previous works, we presented the Multimodal Intrinsic Speckle-Tracking (MIST) algorithm, which recovers the three complementary imaging modes from SBXI data by solving the associated Fokker--Planck equation. In this work, we present a variation of MIST, called ``gradient-flow MIST", which (1) reduces the amount of SBXI data required for image retrieval, (2) maintains the full generality of the X-ray Fokker--Planck equation, and (3) recovers dark-field images with higher quality than the previously proposed variants for weakly attenuating (i.e., low density) samples. We demonstrate the new gradient-flow MIST approach on experimental SBXI data of a knotted bundle of carbon fibres acquired at the Australian synchrotron. This approach is anticipated to be useful in phase-contrast and dark-field applications that require simplicity in experimentation and low sample X-ray exposure.
\end{abstract}

\maketitle

%\tableofcontents

\section{Introduction}
Conventional attenuation-based X-ray imaging (like clinical radiography) provides low image contrast between similarly attenuating structures and yields little information on structures that are below the imaging system's spatial resolution. Phase-contrast imaging (PCI) \citep{10.1063/1.1146073} and dark-field imaging (DFI) \citep{1992RScI...63..611M} are two techniques that address these respective limitations \citep{Pagnin_2006, ENDRIZZI201888, Levine_Long_2004, Moosmann2013, Moosmann_Ershov_Weinhardt_Baumbach_Prasad_LaBonne_Xiao_Kashef_Hofmann_2014}. The basis of attenuation-based imaging has remained unchanged since its discovery in 1896 \citep{Rontgen1895}, relying solely on differences in X-ray attenuation along the X-ray's trajectory to generate contrast. PCI, on the other hand, additionally exploits differences in phase shifts induced by different structures in the sample to generate contrast. The contrast in DFI arises from multiple contrast mechanisms, such as, small angle X-ray scattering from microstructures and Young–Maggi–Rubinowicz boundary-diffraction from edges \citep{Glatter_Kratky_1982,Levine_Long_2004,Miyamoto:62}. These additional channels of information allow PCI and DFI to reveal features that are invisible in attenuation images \citep{Pfeiffer_Weitkamp_Bunk_David_2006, Pfeiffer_Bech_Bunk_Kraft_Eikenberry_Brönnimann_Grünzweig_David_2008}. For example, in medical imaging, phase-retrieved phase-contrast images of the breast show better image quality than their respective attenuation images \citep{Gunaseelan2023, f0dbd53b8e70479faffed8ff590edda7,TavakoliTaba2020,Gureyev:tv5076}, and dark-field images allow pathophysiological changes in the lung to be more easily visualised \citep{velroyen2015grating}.

Currently, a variety of PCI and DFI techniques have been demonstrated, such as analyser-based \citep{Förster_Goetz_Zaumseil_1980, Chapman_Thomlinson_Johnston_Washburn_Pisano_Gmür_Zhong_Menk_Arfelli_Sayers_1997}, propagation-based \citep{Paganin_Mayo_Gureyev_Miller_Wilkins_2002, Gureyev_2020}, grating-based \citep{Momose_Kawamoto_Koyama_Hamaishi_Takai_Suzuki_2003,Pfeiffer_Bech_Bunk_Kraft_Eikenberry_Brönnimann_Grünzweig_David_2008}, edge illumination \citep{Olivo_Arfelli_Cantatore_Longo_Menk_Pani_Prest_Poropat_Rigon_Tromba_2001, PhysRevApplied.19.054042}, and speckle-based X-ray imaging (SBXI) \citep{Morgan_Paganin_Siu_2012, Peverini_2012}. All these methods can combine attenuation imaging, PCI, and DFI; hence, they are multimodal. These multimodal techniques have also been shown to work with non-synchrotron X-ray sources, making them strong candidates for mainstream use, for example, in clinics \citep{Russo_2017}. 

Since its realisation, SBXI has been rising in popularity because it is easy to implement experimentally. SBXI uses a speckle pattern in the near-field regime as an X-ray wavefront marker. An X-ray speckle pattern is generated by inserting a structured, usually random, mask with spatially absorbing features and/or phase-shifting features, such as sandpaper, in a partially coherent X-ray beam \citep{Dainty_1975}. When a sample is placed in the beam downstream of the mask, the reference speckle pattern will be modulated according to the structural features in the sample. By mathematically tracking how speckles in the reference speckle pattern $I_R$ have transformed into the sample speckle pattern $I_S$, the structural and compositional features in the sample can be extracted \citep{PhysRevLett.112.253903}. Different modulations to the reference speckle pattern are attributed to different sample-induced contrast mechanisms: a decrease in intensity reflects that the X-ray beam has been attenuated in that region of the sample, transverse speckle-shifts are due to a phase change of the beam in the sample, and blurring of the speckle pattern--which is also called a local decrease in visibility--is due to dark-field effects \citep{Berujon_Wang_Sawhney_2012, Zdora_Thibault_Zhou_Koch_Romell_Sala_Last_Rau_Zanette_2017, cerbino2008x}.

Initially, speckle modulations were tracked by mathematically comparing $I_R$ and $I_S$ in small regions \citep{Zdora_2018, Celestre2025}. Examples of such methods include X-ray speckle tracking (XST) \citep{Morgan_Paganin_Siu_2012, Peverini_2012}, X-ray speckle scanning (XSS) \citep{Peverini_2012} or unified modulated pattern analysis (UMPA) \citep{Zdora_Thibault_Zhou_Koch_Romell_Sala_Last_Rau_Zanette_2017}.
In 2018, another paradigm of speckle tracking was realised and is called the optical--flow method \citep{Paganin_Labriet_Brun_Berujon_2018}, which employed an equation similar in spirit to the widely adopted transport-of-intensity equation (TIE) \citep{Teague_1983} to model the evolution of the reference speckles in $I_R$ into the sample-modified speckles in $I_S$. In comparison to the aforementioned XST, XSS, or UMPA, the speckles are tracked globally, instead of locally, through the entire image, by solving this TIE--like equation. The initial optical--flow method presented in \citet{Paganin_Labriet_Brun_Berujon_2018} was limited to the case of phase-objects, i.e., attenuation and dark-field effects were neglected. An extension, named `Multimodal Intrinsic Speckle Tracking' (MIST), was realised in 2020 in \citet{Pavlov_Paganin_Li_Berujon_Rougé-Labriet_Brun_2020}
which utilises solving the more general X-ray-Fokker--Planck equation \citep{Paganin_Morgan_2019}. The X-ray Fokker--Planck equation extends the TIE by also including dark-field effects. The inclusion of dark-field effects allows the reconstruction of an image that reveals information about unresolved microstructures, while also improving the quality of the phase-shift image \citep{Alloo_2025}. 
MIST therefore reconstructs three complementary images: attenuation, phase-contrast, and dark-field \citep{Alloo_Paganin_Morgan_Kitchen_Stevenson_Mayo_Li_Kennedy_Maksimenko_Bowden2022}. 

Since its introduction in 2020 by \citet{Pavlov_Paganin_Li_Berujon_Rougé-Labriet_Brun_2020}, several variants of the MIST algorithm have been proposed \citep{Alloo_Paganin_Morgan_Kitchen_Stevenson_Mayo_Li_Kennedy_Maksimenko_Bowden2022,Pavlov_Paganin_Morgan_Li_Berujon_Quénot_Brun_2021,Alloo_Morgan_Paganin_Pavlov_2023} to improve the stability of the solutions and increase the generality of the retrieval approach. Each variant differs in its approach to solving the X-ray Fokker--Planck equation and requires at least two or more speckle image pairs. A speckle image pair consists of the images $I_R$ and $I_S$; different pairs are obtained by shifting the speckle-generating mask and collecting $I_R$ and $I_S$ images at each mask position. The first variant of MIST \citep{Pavlov_Paganin_Li_Berujon_Rougé-Labriet_Brun_2020} assumes the sample is a) pure-phase and b) the dark-field is spatially slow-varying, to simplify solving the X-ray Fokker--Planck equation. Assumption a) was relaxed in \citet{Alloo_Paganin_Morgan_Kitchen_Stevenson_Mayo_Li_Kennedy_Maksimenko_Bowden2022}, as well as extending MIST to tomography. Assumption b) may lead to artefacts at sharp interfaces in the reconstructed dark-field images and was later relaxed in \citet{Alloo_Morgan_Paganin_Pavlov_2023}. The most recent, superior in terms of image quality, and most general MIST variant for isotropic dark-field \citep{Alloo_Morgan_Paganin_Pavlov_2023} requires a minimum of four speckle image pairs, which is an increase from the original variant by a factor of two \citep{Pavlov_Paganin_Li_Berujon_Rougé-Labriet_Brun_2020}. This may limit its applicability in scenarios where image acquisition time should be minimised, such as medical imaging.

In this paper, we present a new MIST variant called gradient-flow MIST (GF-MIST). GF-MIST requires only two speckle image pairs, is robust to noise, and maintains the same generality as \citep{Alloo_Morgan_Paganin_Pavlov_2023}, making it applicable to a broad range of samples. We derive GF-MIST by solving the Fokker--Planck equation for SBXI, and then demonstrate the method's effectiveness on SBXI data of a sample of a knotted carbon fibre collected at the Australian synchrotron.
\begin{figure}[!tb]
	\centering
	\includegraphics[width=0.35\textwidth]{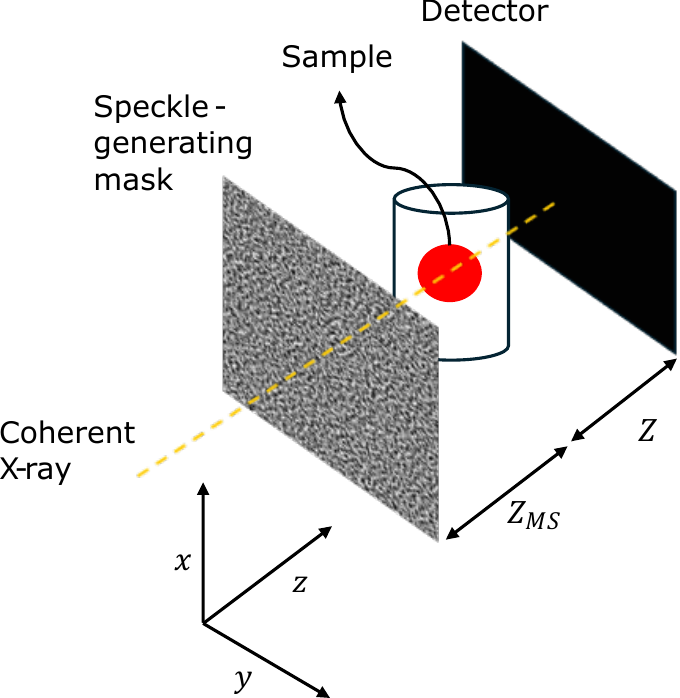}
	\caption{Schematic of a speckle-based X-ray imaging (SBXI) setup, where $Z_{MS}$ is the mask-sample distance, and $Z$ is the sample-detector distance.}  \label{setup}
\end{figure}
\section{Theory} \label{theory}
Consider the SBXI setup for monochromatic paraxial illumination as shown in Figure~\ref{setup}. The evolution of the reference-speckle pattern $I_{R}$ into the sample-speckle pattern $I_{S}$ can be modelled by the finite-difference form of the X-ray Fokker--Planck equation \citep{Alloo_Paganin_Morgan_Kitchen_Stevenson_Mayo_Li_Kennedy_Maksimenko_Bowden2022}:
\begin{align} \label{fokkerplanck}
	I_{R}(\textbf{r}_\perp,Z)I_\text{ob}(\textbf{r}_\perp,0)-I_{S}(\textbf{r}_\perp,Z) =  \nonumber \\ \nabla_{\perp} \cdot \left[\frac{Z}{k}I_{R}(\textbf{r}_\perp,Z)I_\text{ob}(\textbf{r}_\perp,0)\nabla_{\perp}\phi(\textbf{r}_\perp,0)\right] -\nonumber \\
    Z^2\nabla_{\perp}^2\left[D(\textbf{r}_\perp,0)I_{R}(\textbf{r}_\perp,Z)I_\text{ob}(\textbf{r}_\perp,0)\right]. 
\end{align}
Above, $\textbf{r}_\perp\equiv (x,y)$ denotes the Cartesian coordinates perpendicular to the optical $z$ axis, $\nabla_{\perp} \equiv (\partial_x,\partial_y)$ is the 2-D transverse gradient operator, $k$ is the wavenumber given by $2\pi/\lambda$, where $\lambda$ is the wavelength, $I_\text{ob}(\textbf{r}_\perp,0)$ is the sample's transmission at the sample's exit-plane $z=0$,
$I_R(\textbf{r}_\perp,Z)$ and $I_S(\textbf{r}_\perp,Z)$ are the reference and sample speckle images, respectively, collected at a distance $z=Z$ from the sample, $\phi(\textbf{r}_\perp,0)$ is the sample-induced phase-shift, and $D(\textbf{r}_\perp,0)$ is the dimensionless effective diffusion coefficient which characterises local isotropic dark-field effects. Equation~\ref{fokkerplanck} is the SBXI Fokker--Planck equation, accounting for contributions from attenuation, phase-shift, and dark-field effects. This equation 
can be rearranged into a form that is more suggestive of a continuity equation, 
\begin{equation} \label{continuity}
	\frac{I_{R}I_\text{ob}-I_{S}}{Z} -\nabla_{\perp} \cdot \left[\frac{1}{k}I_{R}I_\text{ob}\nabla_{\perp}\phi-Z\nabla_{\perp}\left(DI_{R}I_\text{ob}\right)\right]=0,
\end{equation} 
where the functional dependencies are omitted for clear presentation. The conserved quantity in this continuity equation ~\eqref{continuity} is the integral of the quantity $I_RI_\text{ob}-I_S$ over a 2-D region (energy), and the conserved current $\textbf{J}_\perp(\textbf{r}_\perp,0)$ is given by the term inside the divergence operator:
\begin{equation} 
	\textbf{J}_{\perp}=\frac{1}{k}I_{R}I_\text{ob}\nabla_{\perp}\phi-Z\nabla_{\perp}\left(DI_{R}I_\text{ob}\right). 
\end{equation}
The current $\textbf{J}_\perp$ describes the flow of energy in the transverse direction as $I_RI_\text{ob}$ is transformed into $I_S$. This current term comprises two parts that govern the energy flow: a coherent and a diffusive transport part \citep{Paganin_Morgan_2019}. These two flow mechanisms together give the forward model describing the evolution of $I_R$ into $I_S$, and by solving the associated inverse problem, the attenuation, phase-shift, and dark-field can be recovered. 

To begin solving the inverse problem, that is retrieving $\phi$ and $D$, we take the 2-D Helmholtz decomposition of $\textbf{J}_{\perp}$ into an irrotational and rotational part, characterised by $V(\textbf{r}_\perp,0)$ and $A(\textbf{r}_\perp,0)$, respectively (cf. \citet{Teague_1983}), namely \citep{Schmalz_Gureyev_Paganin_Pavlov_2011}
\begin{equation} \label{Helmholtz}
	\frac{1}{k}I_{R}I_\text{ob}\nabla_{\perp}\phi-Z\nabla_{\perp}\left(DI_\text{ob}I_{R}\right) =\nabla_{\perp} V +  \textbf{rot} (A).
\end{equation}
Above, the rotation operator $\textbf{rot}$ maps $A$ into the vector $(\partial_yA,-\partial_xA)$ \citep{Schmalz_Gureyev_Paganin_Pavlov_2011}. In Eq.~\eqref{Helmholtz}, we neglect the rotational component $\textbf{rot} (A)$, which is a valid assumption if the sample's attenuation does not vary rapidly \citep{Schmalz_Gureyev_Paganin_Pavlov_2011}.
By applying the 2-D divergence to both sides of Eq.~\eqref{Helmholtz} and comparing with Eq.~\eqref{continuity}, the function $V$, which is related to the phase-shift and dark-field images we aim to recover, can be obtained by numerically solving the following Poisson equation:
\begin{align} \label{PoissonV}
\frac{I_RI_\text{ob} - I_S}{Z} = \nabla_{\perp}^2 V.
\end{align}
Equation~\eqref{PoissonV} must be solved subject to appropriate boundary conditions. If the sample lies within the entire field of view of the detector, one can impose fixed Dirichlet boundary conditions \citep{Paganin_Pelliccia_Morgan_2023} as there is no phase-shift or dark-field outside the sample, and hence, at the boundary of the images.
The transmission image $I_\text{ob}$ required to solve Eq.~\eqref{PoissonV} can be obtained by acquiring a contact image, i.e., an attenuation-only image, during imaging (i.e., $Z = 0$). If the sample is approximately single-material, a first order approximation for $I_\text{ob}$ from a phase-retrieval solution derived from the TIE for SBXI can be used \citep{Pavlov_Li_Paganin_Berujon_Rougé-Labriet_Brun_2020}:
\begin{equation} \label{Pavlov}
    I_\text{ob}^{(1)} \approx \mathcal{F}^{-1}\frac{1}{1+\frac{\gamma Z}{2k}\textbf{k}_\perp^2}\mathcal{F}\left(\frac{I_S}{I_R}\right),
\end{equation}
 where $\mathcal{F}$ is the 2-D Fourier transform, $\textbf{k}_{\perp}$ is the 2-D spatial frequency vector, and  $\gamma$ (known \textit{a priori}) is the ratio of the real part of the sample's refractive index to its imaginary, $\gamma = \delta/\beta$. Note that the refractive index, $n$, is given by $n=1-\delta+i\beta$, where $\delta$ and $\beta$ characterise the sample's refraction and attenuation properties, respectively \citep{Pagnin_2006}. We label $I_\text{ob}^{(1)}$ with a superscript `1' to indicate that it is a first-order approximation derived using the TIE and does not account for dark-field effects. An iterative procedure that reintroduces these image-blur-inducing dark-field effects into the solution for $I_\text{ob}$ is described later, thereby improving the transmission retrieval. 
 
Once $V$ is retrieved, it can be used to establish the following set of linear equations,
\begin{align}
	I_{R}\alpha_1+Z(\partial_xI_{R})\alpha_3 \approx\partial_x V,  \label{linearequation1}\\
		I_{R}\alpha_2+Z(\partial_yI_{R})\alpha_3 \approx\partial_y V, \label{linearequation2}
\end{align}
in which the unknown variables are linked to the phase-shift and dark-field images to be retrieved via $\alpha_{1}$--$\alpha_{3}$, which are defined as
\begin{align}
\alpha_{1} &= \frac{1}{k}I_\text{ob}\partial_x\phi-z \partial_x(DI_\text{ob}), \label{a1} \\
\alpha_2 &= \frac{1}
{k}I_\text{ob}\partial_y\phi-z \partial_y(DI_\text{ob}), \label{a2} \\
\alpha_3 &=  -DI_\text{ob}. \label{a3}
\end{align}
\begin{figure*}[!htb]
  \centering
  \includegraphics[width=1\textwidth]{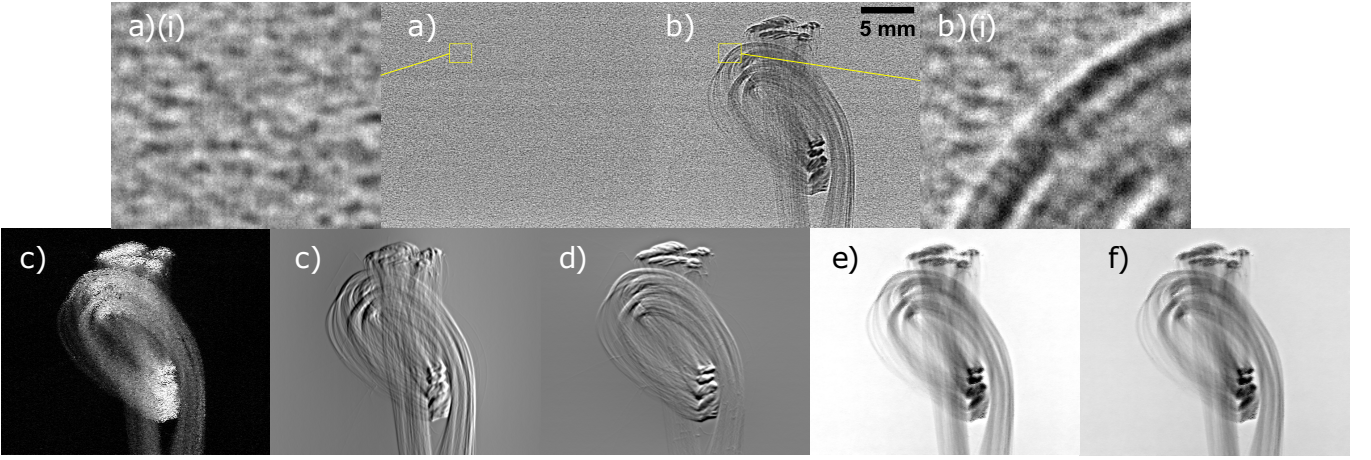}
  \caption{a)--b) SBXI data collected at the Australian synchrotron's Imaging and Medical beamline (IMBL). c)--g) Reconstructed images of a knotted carbon fibre using twenty speckle image pairs with GF-MIST. a) Reference speckle pattern $I_R$, b) sample speckle pattern $I_S$, c) effective diffusion coefficient $D$, d) angle of refraction (corrected for SAXS) in the $x$ direction $\alpha_1$, e) angle of refraction (corrected for SAXS) in the $y$ direction $\alpha_2$, f) transmission $I_\text{ob}^{(2)}$, g) phase-shift $\phi$. The linear greyscale range in [min(black), max(white)] of a)--b) is [0.65, 0.97], c) [0, 78]$\times$$10^{-12}$, d) and e) are [-2.4, 2.4]$\times$$10^{-6}$ radians, f) is [0.87, 1.0], g) [-25, 185] radians.}
  \label{carbonknotreconstruction}
\end{figure*}
The variables $\alpha_1$ and $\alpha_2$ are interpreted as being associated with the refraction angles in the $x$ and $y$ directions, respectively, which consider dark-field effects. The third variable, $\alpha_3$, is a thickness-dependent dark-field term. 
Numerically, $\nabla_\perp V$ can be computed on a square array of pixels with pixel size $\mathcal{W}\times\mathcal{W}$ using Fourier filtering,
\begin{equation}
\nabla_\perp V = \mathcal{F}^{-1}f_df_L\mathcal{F}\left[\frac{I_RI_\text{ob} - I_S}{Z}\right]
\end{equation}
where $f_d$ is the noise-robust derivative filter \citep{Fornberg_1987},
\begin{equation}
f_d = \left(\frac{i\text{sin}(k_x\mathcal{W})}{\mathcal{W}},\frac{i\text{sin}(k_y\mathcal{W})}{\mathcal{W}}\right),
\end{equation}
and $f_L$ is the inverse $\nabla_\perp^2$ filter \citep{Paganin_Favre-Nicolin_Mirone_Rack_Vilanova_Olbinado_Fernandez_daSilva_Pelliccia_2020,Press2007},
\begin{equation}
f_L = \frac{\mathcal{W}^2}{2\text{cos}(k_x\mathcal{W})+2\text{cos}(k_y\mathcal{W})-4}.
\end{equation}
The filter $f_L$ is numerically unstable at the origin due to division by zero. However, as we only require Laplacian inversion to recover the gradient of $V$, we can simply set the solution of the inversion to zero at the origin. This corresponds to shifting $V$ by a constant value, which is possible since the gradient of $V$ gives the same solution as an un-shifted variant. Equations~\eqref{linearequation1}--\eqref{linearequation2} are equivalent to Eq.~\eqref{Helmholtz} but with the orthogonal $x$ and $y$ components forming one equation each, and the rotational part (i.e, $\textbf{rot}(A)$) neglected. This linear system is underdetermined, i.e., it contains more unknowns than equations, and therefore no unique solution exists. An approach applied in all MIST algorithms is to construct additional sets of equations by using multiple sets of speckle image pairs. A solution that minimises the residuals of this overdetermined system can be computed using a linear least squares approach \citep{Alloo_Morgan_Paganin_Pavlov_2023}, such as Tikhonov regularised-QR decomposition \citep{tikhonov1977solutions}. However, if only one set of equations is available, then the system is underdetermined.
\begin{figure}[!htb]
  \centering
  \includegraphics[width=0.42\textwidth]{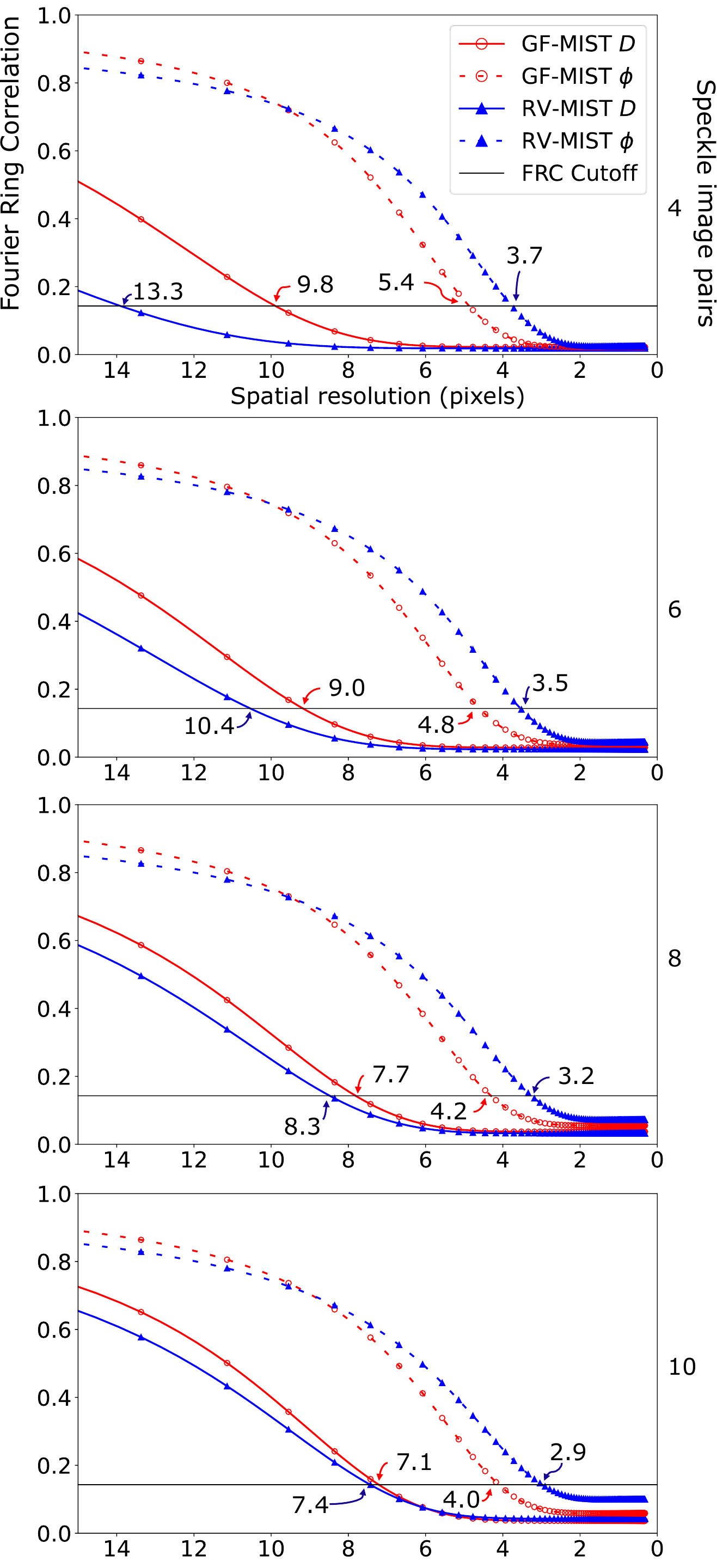}
  \caption{Fitted Fourier ring correlation (FRC) curves for the retrieved dark-field and phase-shift images (solid and stripped lines, respectively) in Figure~\ref{carbonknotcomparison}. The FRC's were fitted to a complementary error function. The spatial resolution was defined as the intersection (labelled) between the FRC curves and a threshold cutoff at 0.143, denoted by the solid horizontal line.}
  \label{FRC}
\end{figure}

When Eq.~\eqref{Pavlov} is used to approximate $I_\text{ob}$, an extra step may be required to correct for dark-field effects if the sample is strongly scattering, as the TIE-retrieved transmission image will be significantly blurred in these areas. To consider dark-field effects in the attenuation retrieval step, Eq.~\eqref{Pavlov} should be extended to include the diffusion term in the X-ray Fokker--Planck equation. This is achieved by expressing a closed-form solution to $I_\text{ob}$ from Eq.~\eqref{continuity} instead of from the TIE. Making use of the projection approximation, $\phi=-k\delta T$, where $T$ is the projected thickness of the sample, and Beer's law of absorption, $I_\text{ob}=\text{exp}(-2k\beta T)$ \citep{Pagnin_2006, Paganin_Nugent_1998, Pavlov_Li_Paganin_Berujon_Rougé-Labriet_Brun_2020}, we rewrite Eq.~\eqref{continuity} as
\begin{equation} \label{rearranged}
\left(1-\frac{\gamma Z}{2k}\nabla^2_{\perp}\right)I_\text{ob} = g_1+g_2
\end{equation}
where we define the source terms
\begin{equation}
g_1(\textbf{r}_{\perp}) \equiv \frac{I_S}{I_R},
\end{equation}
\begin{align}
g_2(\textbf{r}_{\perp}) \equiv \frac{Z}{I_R}\left(\alpha_1\partial_xI_R+\alpha_2\partial_yI_R\right. \nonumber \\
-Z\left.[I_R\nabla_\perp^2\alpha_3+\alpha_3\nabla_\perp^2I_R+\nabla_\perp \alpha_3\cdot\nabla_\perp I_R]\right).
\end{align} 
Equation~\eqref{rearranged} has a closed-form solution for the transmission which considers dark-field effects:
\begin{equation} \label{pavlovextend}
    I_\text{ob}^{(2)} = \mathcal{F}^{-1}\frac{1}{1+\frac{\gamma Z}{2k}\textbf{k}_\perp^2}\mathcal{F}\left(g_1+g_2\right).
\end{equation}
\begin{figure*}[!htb]
  \centering
  \includegraphics[width=\textwidth]{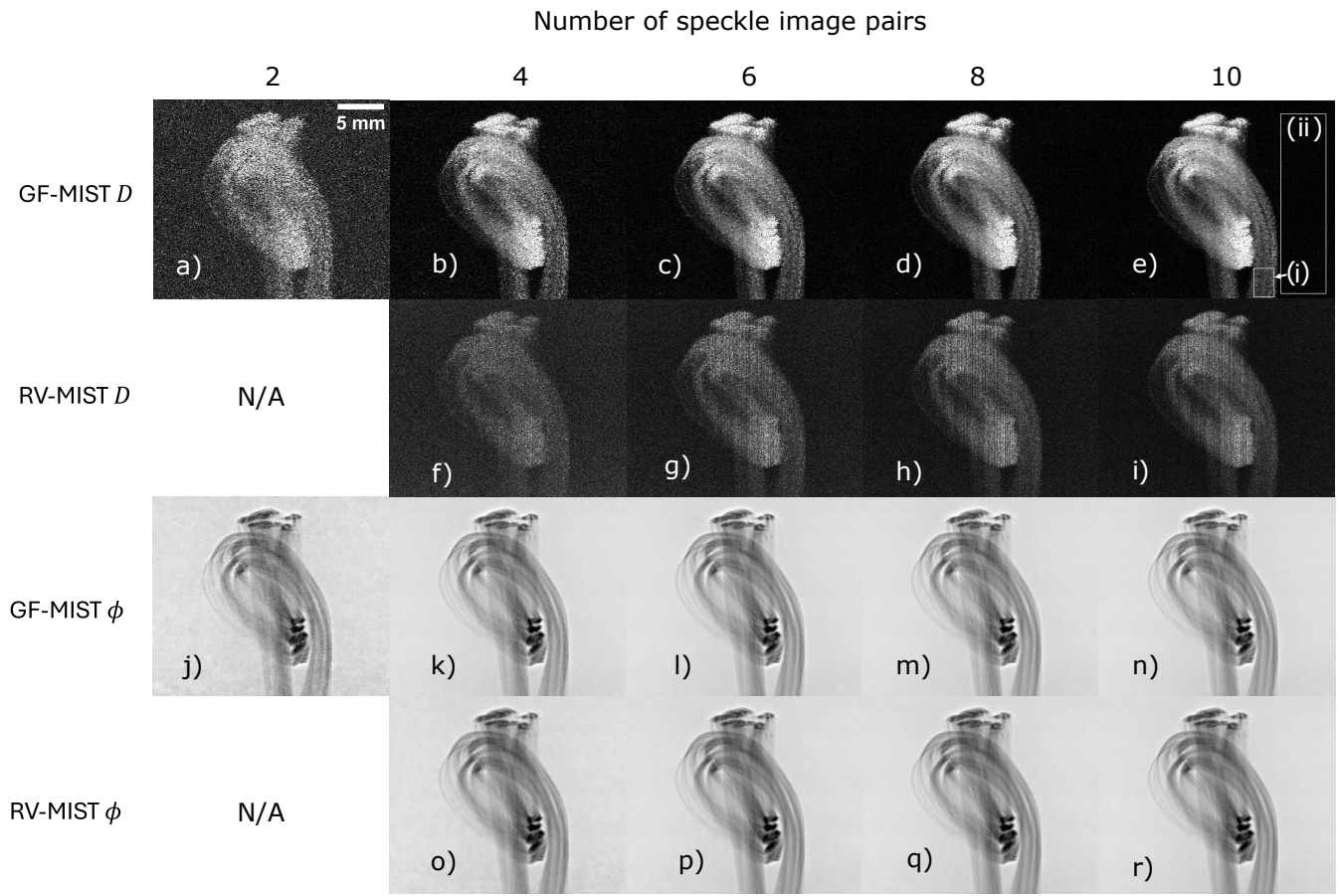}
  \caption{Retrieved dark-field $D$ and phase-shift $\phi$ images for a carbon-knot with GF-MIST (a)-e) and j)-n)) and RV-MIST (f)-i) and o)-r)) using 2, 4, 6, 8, and 10 speckle-image pairs. The linear greyscale range for images a)--e)$=[0,70]\times10^{-12}$, f)--i)$=[0,30]\times10^{-12}$, j)--r)$=[-185,25]$ radians. The white box in e)(i) and e)(ii) indicates the signal and background area used for the calculation of CNR for images a)--r), respectively.}
  \label{carbonknotcomparison}
\end{figure*}
Here, we have separated the dominant term, $g_1$, from $g_2$, which represents small second-order correction terms to $g_1$. The unknown variables $\alpha_1$, $\alpha_2$, and $\alpha_3$ used in the right-hand side of Eq.~\eqref{pavlovextend} are computed with $I_\text{ob}^{(1)}$ first.
The diffusion coefficient dark-field can then be retrieved directly using the solution $\alpha_3$ and the updated transmission $I_\text{ob}^{(2)}$, that is,
\begin{equation} \label{D}
D = -\frac{\alpha_3}{I^{(2)}_\text{ob}}.
\end{equation}
The phase-shift can then be retrieved using 
\begin{equation}\label{phi}
    \phi = \frac{\gamma}{2}\text{ln}\left[I_\text{ob}^{(2)}\right].
\end{equation}
In the case when a contact image is used to obtain an exact value of $I_\text{ob}$, then dark-field is retrieved from 
\begin{equation} \label{D2}
D = -\frac{\alpha_3}{I_\text{ob}}, 
\end{equation}
and the phase-shift is found by numerically solving the following differential equation:
\begin{equation} \label{phaseequation}
I_\text{ob}-\frac{Z}{k}\left[\nabla \cdot I_\text{ob}\nabla \phi\right]= g_1+g_2,
\end{equation}
 for example, using finite difference methods \citep{Press2007}. Equation~\eqref{phaseequation} was formed by expressing Eq~\eqref{continuity} in terms of $g_1$ and $g_2$.
\section{Methods}
To validate our approach, SBXI data of a knotted carbon fibre was collected at the Australian synchrotron's Imaging and Medical beamline (IMBL) using a setup similar to that shown in Fig.~\ref{setup}. The knotted carbon fibre was placed $Z=3$ m from the detector and $Z_{MS} = 1$ m from the speckle-generating mask. The mask consisted of a stack of six sandpapers with grit size P240. X-rays of 25 keV with an energy bandwidth of $\Delta E/E=10^{-3}$ were used to illuminate the mask and the sample. Twenty sets of speckle image pairs were collected by transversely shifting the sandpaper. The effective pixel size for the detector was 10.2 $\mu$m, while the speckle pattern had an effective speckle size of 105.9 $\mu$m, which was computed by fitting the autocorrelation of $I_R$ to a Gaussian function and taking the full-width-half-maximum \citep{goodman2007speckle, 88d937a3476a42f6b9b5490418b4da06}. 

\subsection{Numerical Implementation}
Our developed GF-MIST algorithm was implemented in MATLAB (version 2023b), with the script and test data openly available at \cite{GF-MIST}. For this carbon fibre sample, we assumed it to be single-material under 25 keV X-ray illumination and therefore had a $\gamma$ value of $2600$, which was computed from the TS imaging calculator \citep{AAA}. The transmission was then initially approximated using the TIE-derived solution (Eq.~\eqref{Pavlov}). The overdetermined system of linear equations generated by twenty instances of Eqs.~\eqref{linearequation1}-\eqref{linearequation2} (totalling 40) was solved with Tikhonov-regularised QR decomposition using a Tikhonov parameter of $4.6\times10^{-5}$, determined using the L-curve method \citep{Hansen_O’Leary_1993}. To then consider dark-field effects, thereby correcting for scatter-induced image blur, the transmission was updated using Eq.~\eqref{pavlovextend}. The dark-field and phase-shift were then extracted using Eqs.~\eqref{D} and \eqref{phi} using the updated transmission.

\subsection{Evaluation of Image Quality}
We evaluated GF-MIST's performance by comparing its retrieved images to those obtained using the MIST approach developed in \citep{Alloo_Morgan_Paganin_Pavlov_2023}, which we call `RV-MIST' hereafter. We used the openly available Python script at \citep{Alloo2023MIST} to obtain the RV-MIST reconstructions. 

The Fourier ring correlation (FRC) was used to quantify the spatial resolution of the images, that is, the smallest length scale resolvable in the images. The FRC measures the correlation of two images in the frequency domain and is defined as 
\begin{equation}
    \text{FRC} = \frac{\sum_{|\textbf{k}_\perp|} F_1(\textbf{k}_\perp) \cdot F_2^{*}(\textbf{k}_\perp)}{\sqrt{\sum_{|\textbf{k}_\perp|}|F_1(\textbf{k}_\perp)|^2\cdot\sum_{|\textbf{k}_\perp|}|F_2(\textbf{k}_\perp)|^2}}
\end{equation}
where $F_1(\textbf{k}_\perp)$ and $F_2(\textbf{k}_\perp)$ are the Fourier transforms of image 1 and image 2, respectively, of the same sample. Images 1 and 2 were obtained by splitting the full twenty SBXI dataset in half and applying the GF-MIST/RV-MIST to each half.  
As the FRC curves were noisy, they were fitted to a complementary error function (erfc). The 0.143 FRC threshold criterion \citep{VANHEEL2005250} was then used as a cutoff for the spatial resolution. 

The contrast-to-noise ratio (CNR) was used to measure the visibility of structures within the images, which was computed using
\begin{equation}
\text{CNR}=\frac{|\mu_\text{s}-\mu_\text{b}|}{\sigma_\text{s}}
\end{equation}
where $\mu_\text{s}$ and $\sigma_\text{s}$ are the mean and standard deviation of the intensity of an area of a flat signal, respectively, and $\mu_\text{b}$ is the mean background intensity.
\section{Results and Discussion}
Figure~\ref{carbonknotreconstruction} shows the GF-MIST-retrieved dark-field, refraction angles in the transverse directions, phase-shift, and transmission images reconstructed using the entirety of the SBXI of the carbon knot, i.e., 20 speckle image pairs. Notably, GF-MIST is the first Fokker--Planck-equation-derived algorithm for SBXI to retain the speckle-shift term $\nabla I_R \cdot \nabla\phi$ \citep{Pavlov_Li_Paganin_Berujon_Rougé-Labriet_Brun_2020}, and hence can reconstruct the two orthogonal refraction angles (i.e., consider phase gradients within the sample) without requiring to calculate the phase-shift first. Due to their main orientation, the fibres are the most visible in the reconstruction of the $x$ refraction angle. However, they can also be seen on the $y$ refraction angle, phase-shift, transmission, and partially on the dark field images. 

Figure~\ref{carbonknotcomparison} compares dark-field and phase-shift images retrieved using GF-MIST and RV-MIST, and shows how the retrieved images are affected by the number of input speckle image pairs. RV-MIST requires at least four speckle image pairs, whereas GF-MIST needs a minimum of two. The magnitude of the GF-MIST's retrieved dark-field signal is $\sim2$ times that of that obtained using RV-MIST, and hence, the maximum and minimum displayed greyscale values in Fig.~\ref{carbonknotcomparison} had to be set differently for each to allow for appropriate qualitative comparison. The difference in the retrieved magnitude of the dark-field signal may arise from the fact that GF-MIST computes first derivatives of $I_R$ when constructing its system of linear equations, whereas RV-MIST computes the second derivative (i.e., transverse Laplacian operator) of $I_R$. This reduction in one derivative order results in less high-frequency noise amplification, allowing more dark-field signal to be retrieved through the noise. In other words, from the continuous Fourier derivative theorem, the first derivative of $I_R$ is equivalent to linear scaling by $i(k_x,k_y)$ in the frequency domain, whereas the second derivative is quadratically scaled by $\textbf{k}_{\perp}^2$, which is less numerically robust to noise. For both methods, however, the phase-shifts quantitatively agree. There also appear to be vertical stripe artefacts that are reconstructed in the dark-field images of RV-MIST, but not in GF-MIST. 

To compare the image quality of the dark-field and phase-shift images retrieved using RV-MIST and GF-MIST across different numbers of speckle image pairs, the spatial 
\begin{figure}[!htb]
  \centering
  \includegraphics[width=0.50\textwidth]{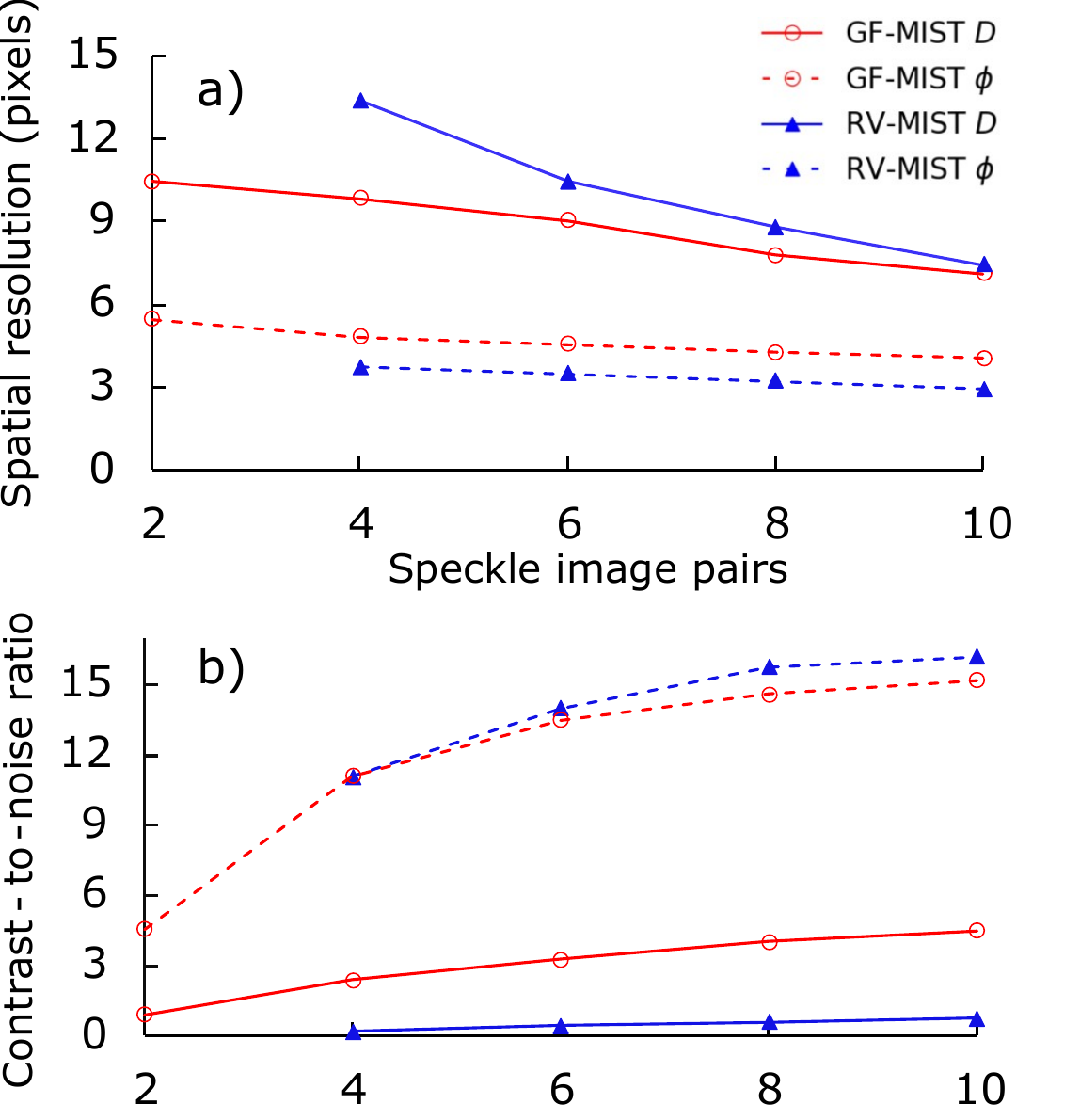}
  \caption{a) Dark-field (solid line) and phase-shift (dashed line) spatial resolution of GF-MIST (circles) and RV-MIST (triangles). b) same as in a) but for the measured contrast-to-noise ratio (CNR) using the signal and background as shown in Figs. \hyperref[carbonknotcomparison]{\ref*{carbonknotcomparison}(e)(i)} and \hyperref[carbonknotcomparison]{\ref*{carbonknotcomparison}(e)(ii)}, respectively.}
  \label{comparisongraph}
\end{figure}
resolution and CNR were computed. Figure~\ref{FRC} shows FRC curves for both approaches. The general trend for both approaches is that increasing the number of speckle image pairs shifts the FRC curve to the right, indicating improved spatial resolution. 

Figure~\hyperref[comparisongraph]{\ref*{comparisongraph}a} plots the FRC-computed spatial resolution, as a function of speckle image pairs used to reconstruct the dark-field and phase-shift images using GF-MIST and RV-MIST. Figure~\hyperref[comparisongraph]{\ref*{comparisongraph}b} does the same but with the computed CNR. The dark field spatial resolution of RV-MIST approximately lags behind GF-MIST by two speckle image pairs. For example, the dark field image retrieved from GF-MIST with four pairs yields comparable resolution to that obtained from RV-MIST using six. This is particularly noteworthy, as it suggests that comparable image quality can be achieved with reduced sample X-ray exposure.

Regarding the phase shift, the resolution of GF-MIST is approximately one pixel lower than that of RV-MIST. This is because computing terms such as $\alpha_3\nabla_\perp^2I_R$ and $\nabla_\perp\alpha_3\cdot\nabla_\perp  I_R$ in the correction factor $g_2$ (Eq.~\eqref{pavlovextend}), for reasons mentioned previously, amplifies noise. For the same reasons, the CNR for the GF-MIST-retrieved dark-field images is up to $\sim10$ times greater than RV-MIST using the same number of speckle imaging pairs, and slightly below RV-MIST for the phase-shift images.  

While we demonstrate GF-MIST's superior dark-field performance over RV-MIST in this manuscript, this result corresponds to the ideal case for GF-MIST--namely, a low-attenuating sample. Evidently, there will be cases in which RV-MIST outperforms GF-MIST, and even cases where earlier versions of MIST, such as those proposed by \citet{Pavlov_Paganin_Li_Berujon_Rougé-Labriet_Brun_2020} and \citet{Alloo_Paganin_Morgan_Kitchen_Stevenson_Mayo_Li_Kennedy_Maksimenko_Bowden2022}, may outperform both RV- and GF-MIST. Although earlier variants of MIST rely on stronger assumptions, they are mathematically simpler to solve, which can affect the stability and, consequently, the quality of the retrieved images. It would be valuable for future work to investigate the performance of RV-MIST, GF-MIST, and earlier variants across different types of samples, including highly attenuating specimens or those with hierarchical structures that vary rapidly in space.

\section{conclusion}
We presented GF-MIST, a general and noise-robust MIST algorithm for SBXI, which uses a gradient-flow approach to solving the X-ray Fokker--Planck equation. This approach was applied to speckle data of a knotted carbon fibre to successfully recover the attenuation, phase-shift and dark-field images, in addition to the refraction angles from the sample. We compared the image quality of the dark-field and phase-shift images reconstructed from GF-MIST to the most recent MIST method, which we call `RV-MIST' \citep{Alloo_Morgan_Paganin_Pavlov_2023}. GF-MIST reconstructs dark field images at higher quality, at the cost of a slight decrease in the phase-shift image quality. GF-MIST also requires fewer speckle image pairs than RV-MIST to operate; therefore is suitable for applications where experimental simplicity and image acquisition speed are required. 
\begin{acknowledgments}
This research was funded by the French Agence Nationale de la Recherche and France 2030, grants MIAI Cluster (ANR-23-IACL-0006) and Labex CAMI (ANR-11-LABX-0004).
Samantha J. Alloo acknowledges funding support from an AINSE Ltd. Early Career Researcher Grant (ECRG). We acknowledge the beamline scientists at the Imaging and Medical Beamline at the Australian Synchrotron for their support in acquiring the imaging data. 
\end{acknowledgments}
\newpage
\bibliography{bibliography}% Produces the bibliography via BibTeX.

\end{document}